\documentclass[useAMS,usenatbib,psfig]{mn2e}
\usepackage{graphicx,rotating,times}
\usepackage{longtable}
\usepackage{amsmath,amssymb,rotating}
\usepackage{lscape}
\def \cm2{\mbox{cm$^{-2}$}}
\def \cm3{\mbox{cm$^{-3}$}}

\title[Total Eclipse of the Heart]{Heart of Darkness: dust obscuration of the central stellar component in globular clusters younger than $\sim$100\,Myr in multiple stellar population models}

\author[S.~N.~Longmore]{S. N. Longmore$^{1}$\thanks{E-mail:~s.n.longmore@ljmu.ac.uk} \\
$^{1}$ Astrophysics Research Institute, Liverpool John Moores University, 146 Brownlow Hill, Liverpool L3 5RF, UK\\
} 
\begin{document}

\date{Accepted MNRAS Letters}
\pagerange{\pageref{firstpage}--\pageref{lastpage}} \pubyear{2015}
\maketitle

\begin{abstract}

To explain the observed anomalies in stellar populations within globular clusters, many globular cluster formation theories require two independent episodes of star formation. A fundamental prediction of these models is that the clusters must accumulate large gas reservoirs as the raw material to form the second stellar generation. We show that young clusters containing the required gas reservoir should exhibit the following observational signatures: (i) a dip in the measured luminosity profile or an increase in measured reddening towards the cluster centre, with A$_V>10$\,mag within a radius of a few pc; (ii) bright (sub)mm emission from dust grains; (iii) bright molecular line emission once the gas is dense enough to begin forming stars. Unless the IMF is anomalously skewed towards low-mass stars, the clusters should also show obvious signs of star formation via optical emission lines (e.g. H$\alpha$) after the stars have formed. These observational signatures should be readily observable towards any compact clusters (radii of a few pc) in the nearby Universe with masses $\gtrsim 10^6$\,M$_\odot$ and ages $\lesssim$100\,Myr. This provides a straightforward way to directly test globular cluster formation models which predict large gas reservoirs are required to form the second stellar generation. The fact that no such observational evidence exists calls into question whether such a mechanism happens regularly for YMCs in galaxies within a few tens of Mpc.

\end{abstract}


%------------------------------------------------------------
\section{Introduction}

Globular clusters (GCs) were once thought to contain stellar populations with the same age, metallicity and chemical abundances. However, it has become clear over the last few decades that this picture is over simplified. When observed with sufficient photometric and spectroscopic precision, all GCs display anomalies, from multiple discrete sequences in colour-magnitude diagrams to variations in the abundances of light elements \citep[e.g.][]{gratton_carretta_bragaglia12}. These anomalies offer an opportunity to constrain the formation history of GCs -- a challenging task given their formation at early epochs of the Universe.

Several GC formation models have been put forward to explain the origin of the observed anomalies \citep[e.g.][]{decressin07, dercole2008, demink09, conroy12, bastian13}. Most of these invoke two or more episodes, or  `generations', of star formation, in which the second generation of stars forms from gas that has been enriched by material from certain subsets of stars in the first generation. A fundamental prediction of the `AGB' model scenarios (in which the polluters are asymptotic giant branch (AGB) stars) is that GCs must accumulate a large gas reservoir after the first generation have finished forming, in order to have sufficient material to form the second generation of stars \citep{dercole2008, dercole2010,conroy_spergel11}. 

In this paper we attempt to constrain the expected properties of this gas reservoir. We then consider how the gas may affect the observed properties of young GCs in the phase between the formation of the first and later generation of stars. As GC formation models make no distinction between the physics of star/cluster formation at the present day and earlier epochs of the Universe (when GCs were forming), these same effects should be seen towards clusters of similar mass, density and age in the local Universe. 

\vspace{-5mm}
%------------------------------------------------------------
\section{Expected properties of gas destined to form the second generation of stars}
\label{sec:gas_models}

%------------------
\subsection{An idealised GC}
\label{sub:toy_model}

We first consider an idealised GC which forms two distinct stellar populations. The first generation of stars has a total initial mass of $M^*_1$, that all form at a time $t_1=0$, i.e. as a single stellar population (SSP). These stars are contained within a radius, R$^*_1$. At some time later, $t^{\rm start}_2$, the second generation of stars begin to form. Star formation proceeds for a time interval of $\Delta t_2$, until the second generation of stars finishes forming at $t^{\rm end}_2 =  t^{\rm start}_2 + \Delta t_2$. The resulting mass of the second generation of stars, $M^*_2$, is a fraction, $F_2$, of the first generation (i.e. $M^*_2 = F_2  M^*_1$). The time-averaged star formation rate required to form the second generation of stars, SFR$^{\rm av}_2$ is then given by SFR$^{\rm av}_2=F_2 M^*_1/\Delta t_2$. R$^*_2$ is defined as the radius containing the second generation of stars at $t^{\rm end}_2$. In order to form a mass $M^*_2$ of stars, a total gas reservoir of mass, $M^{\rm TotGas}_2 = M^*_2/\epsilon $ is required, where $\epsilon$ is the star formation efficiency.  

To match the observed anomalies, the two-generation GC formation models have two distinct gas components -- a component of enriched material expelled from stars within the cluster, and a component of `pristine' (i.e. unprocessed) material. In AGB models, the pristine gas must be accreted from outside the cluster boundary. The mass and spatial distribution of gas within the GC after the first generation of stars have formed must therefore be time variable \citep[see][]{dercole2010}. In order to match the observed light element abundance variations, the two gas components must be at least partially mixed. This implies that both gas components have a high volume-filling factor, i.e. the gas distribution is fairly uniform. We discuss the implications of relaxing this assumption in $\S$\ref{sec:implications}. In the absence of  numerical modelling to study the gas accumulation and star formation process in detail\footnote{We consider more realistic gas profiles in $\S$~\ref{sub:infall_model}},  it seems reasonable to assume that the volume containing the gas before star formation begins (i.e. at time $t^{\rm start}_2$) is similar to the final volume containing the second generation stars\footnote{Observations show that young massive cluster progenitor clouds in the Milky Way are only factors of a few larger than the final stellar populations \citep[][Walker et al. 2015]{longmore14ppvi}}. Following these considerations, we make the simplifying assumptions that: (i) the radius containing the gas, R$_{\rm gas}$, is equal to R$^*_2$; (ii) R$_{\rm gas}$ remains fixed from $t^{\rm start}_2$ to $t^{\rm end}_2$; and (iii) the gas within this radius is uniformly distributed.

The instantaneous gas mass within the cluster, $M^{\rm InstGas}_2$, and the instantaneous star formation rate, SFR$^{\rm Inst}_2$, will vary with  time \citep[see e.g.][]{dercole2010, dercole2011, conroy_spergel11}. In order to calculate the expected range in $M^{\rm InstGas}_2$ and SFR$^{\rm Inst}_2$ over the formation period of the second generation of stars, we consider two extreme formation scenarios. In the `continuous' scenario, the gas accumulation rate is constant between $t^{\rm start}_2$ and $t^{\rm end}_2$. The instantaneous mass accretion rate, $\dot{M}$, is then $\dot{M} = M^{\rm TotGas}_2 /\Delta t = M^*_2 / \epsilon \Delta t$.  The instantaneous star formation rate is constant at a value of SFR$^{\rm Inst}_2 = \epsilon \dot{M}$. In the second `burst' scenario, the full gas reservoir, $M^{\rm TotGas}_2$, is in place at $t^{\rm start}_2$. This gas is then all converted into stars with a star formation efficiency, $\epsilon$, in a single star formation event lasting a time, $\Delta t_2$. We adopt a high value of $\epsilon=0.5$ to provide a conservative lower limit to the required gas reservoir (and hence a lower limit to the extinction and mm-continuum flux density later in the paper).

Having constructed this template, we now consider appropriate values for the parameters defined above. Due to stellar and dynamical evolution over close to a Hubble time, the initial mass ($M^*_1$) and radius (R$^*_1$) of the first stellar generation in GCs are not empirically well constrained. Indeed, inferring $M^*_1$ and $F_2$ from the present-day stellar populations in GCs is model dependent. For example, to solve the `mass budget' problem, \citet{dercole2008} require GCs to have been a factor $\sim10$ more massive at birth than observed today. However, the mass of stars formed in the second generation, $M^*_2$, is constrained much more robustly. Present-day second generation stars are confined to a smaller volume than that of present-day first generation stars \citep[][]{lardo2011}. In order to minimise the mass budget problem, it is generally assumed that the majority of second generation stars have remained bound since forming. In other words, modulo stellar evolution, the measured present-day mass of second generation stars is equal to $M^*_2$. It follows that the present-day mass of second generation stars is also a robust way to estimate the required gas reservoir. 

Although the initial size of GCs is not well constrained, a surprisingly uniform, present-day, half-mass radii of 3\,pc is observed for GCs both in the Milky Way (MW) GCs and in external galaxies \citep{harris96, masters10}. As mentioned above, second generation stars are confined to smaller radii than first generation stars. We therefore adopt an upper limit for R$^*_2 $ of 2\,pc {\citep[see][for observed spatial distributions of first and second generation stars]{lardo2011}}.

Limits for gas accumulation timescales in AGB GC formation models are set by the window in which SNe~II in the first generation have finished and the time at which SNe Ia from the first generation begin. The justification for these limits are that SNe~II clear out any remaining pristine gas that did not end up in stars when the first generation formed, and the SNe Ia can clear any gas that builds up in the GC before star formation can begin. This gives an upper limit on the star formation timescale of $t^{\rm start}_2 \sim 30$\,Myr to $t^{\rm end}_2 \sim 100$\,Myr \citep{dercole2008}\footnote{We note that \citet{conroy_spergel11} adopt a longer (but not directly specified) limit for $t^{\rm end}_2$ of a few 100\,Myr}, which we adopt for the `continuous' star formation scenario. For the `burst' scenario, the upper time limit, $t^{\rm end}_2 \sim 100$\,Myr, is the same. However, in this scenario, while the gas can continue to accumulate to high density, some mechanism must stop it from forming stars. In the model of \citet{conroy_spergel11}, the high Lyman-Werner photon density at cluster ages $\lesssim10^8$\,yr photodissociates molecular hydrogen (H$_2$), thereby suppressing star formation. At any given time it is assumed that the cluster contains a gas reservoir with a mass $\sim$10\% of the stellar mass. This helps the cluster to sweep up pristine gas from the surrounding environment. After $t\sim10^8$\,yr the Lyman-Werner photon density drops precipitously, allowing the gas to catastrophically cool and form stars. In the `burst' scenario, we adopt $\Delta t_2$ to be $\sim$1\,Myr, corresponding to an upper limit of the observed age spread in young massive clusters \citep[cf.][]{longmore14ppvi}. 

%------------------
\subsection{Infall model}
\label{sub:infall_model}

The idealised scenario above is useful to constrain the expected range of gas properties in young GCs. We use the results of \citet{dercole2008} to consider more realistic gas profiles. \citet{dercole2008} constructed a model of gas infalling towards the centre of a GC. This model is ideal for our purposes, as the infalling gas reservoir is the raw material for the second generation of stars. We extracted an approximate density, $\rho$, and temperature, $T$, profile for the fiducial model in their paper, when the GC is at an age of 100\,Myr \citep[][Figure 1]{dercole2008}. From this figure we took the density and temperature structure to be power laws of the following form\footnote{We tried several different parametric fits and the results are not sensitive to the exact form we chose. We therefore opted for the simplest, power law representation.}: $\rho(r) = 851\,(r/{\rm pc})^{-2.2}$\,cm$^{-3}$ and $T(r) = 1548\,(r/{\rm pc}) ^{0.34}$\,K. These reproduce the density and temperature profiles over the range 0.04\,pc~$< r <$~60\,pc. The lower limit to the fit of 0.04\,pc is set by the minimum radius in Figure 1 of \citet{dercole2008}. We then constructed a spherical model of the gas from these profiles and used this to calculate the expected observational effect of this gas. To avoid unphysical temperatures and densities as $r\rightarrow0$, we assumed the temperature and density at radii less than 0.04\,pc to equal those at $r=0.04$\,pc -- i.e. a flat temperature and density profile for $r<0.04$\,pc. In practice, this provides a lower limit to the expected extinction, gas emission and dust emission at $r<0.04$\,pc.

%------------------------------------------------------------
\section{Implications for the observed properties of young GCs}
\label{sec:implications}

We now consider how the gas mass accumulation and subsequent star formation may affect the observable properties of young GCs. As our empirical understanding of GCs is gleaned almost exclusively from observations at optical and infrared wavelengths, we start by considering how gas accumulation and star formation would affect such observations. One potential effect of the gas would be extinction and reddening of the light due to the dust associated with the gas. With an estimate of the gas surface density from $\S$~\ref{sec:gas_models} we can simply use a gas-to-dust ratio and an extinction law to calculate the expected extinction and reddening. To convert from the atomic gas column density, N$_{\rm H}$, to V-band (wavelength of 5470\,\AA) dust extinction, A$_{\rm V}$, we start with a conversion of A$_{\rm V}$/N$_{\rm H} = 5.3 \times 10^{-22}$\,mag\,cm$^2$\,H$^{-1}$, appropriate for Milky Way dust \citep{draine11}. Making the common assumption that the gas-to-dust ratio scales linearly with metallicity \citep[a reasonable assumption for different star formation environments over several orders of magnitude in metallicity: see][]{fisher14}, we can extrapolate the expected extinction to lower metallicity environments.

\begin{figure}
\includegraphics[height=0.48\textwidth, angle=-90]{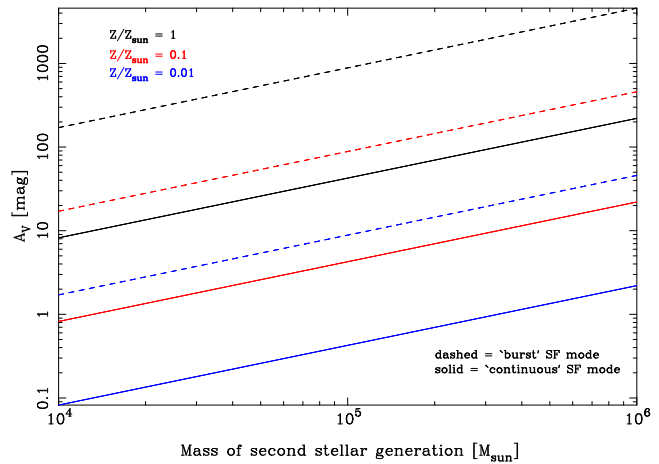}
\caption{Expected V-band extinction as a function of mass in the second generation of stars, $M^*_2$. The black, red and blue lines represent gas-to-dust ratios of 10$^2$, 10$^3$ and 10$^4$, respectively. Assuming the gas-to-dust ratio scales linearly with metallicity, the lines represent metallicities of 1, 0.1, and 0.01 times solar. The solid and dashed lines shows the expected extinction for the `continuous' and `burst' scenarios, respectively}
\label{fig:glob_extinction}
\end{figure}

Figure~\ref{fig:glob_extinction} shows the resulting expected extinction in the V-band as a function of the mass in the second generation of stars (i.e. $M^*_2$ as defined above) in the idealised GC ($\S$~\ref{sub:toy_model} -- i.e. when the gas is confined to a radius of 2\,pc). The black, red and blue lines represent gas-to-dust ratios of 10$^2$, 10$^3$ and 10$^4$, respectively. Assuming the gas-to-dust ratio scales linearly with metallicity, the lines represent metallicities of 1, 0.1, and 0.01 times solar. The solid and dashed lines shows the expected extinction for the `continuous' and `burst' scenarios, respectively. For a given gas-to-dust ratio (or metallicity) and $M^*_2$, the values between the solid and dashed lines show the range of expected extinction. It is clear that for gas-to-dust ratios $\leq10^3$ (i.e. $Z/Z_\odot \geq$ 0.1) the V-band extinction is always $>$1\,mag. As such, the gas and dust would be expected to have a significant effect on the observed optical magnitudes and colours of young GCs. 

\begin{figure*}
\begin{tabular}{cc}
\includegraphics[height=0.48\textwidth, angle=-90]{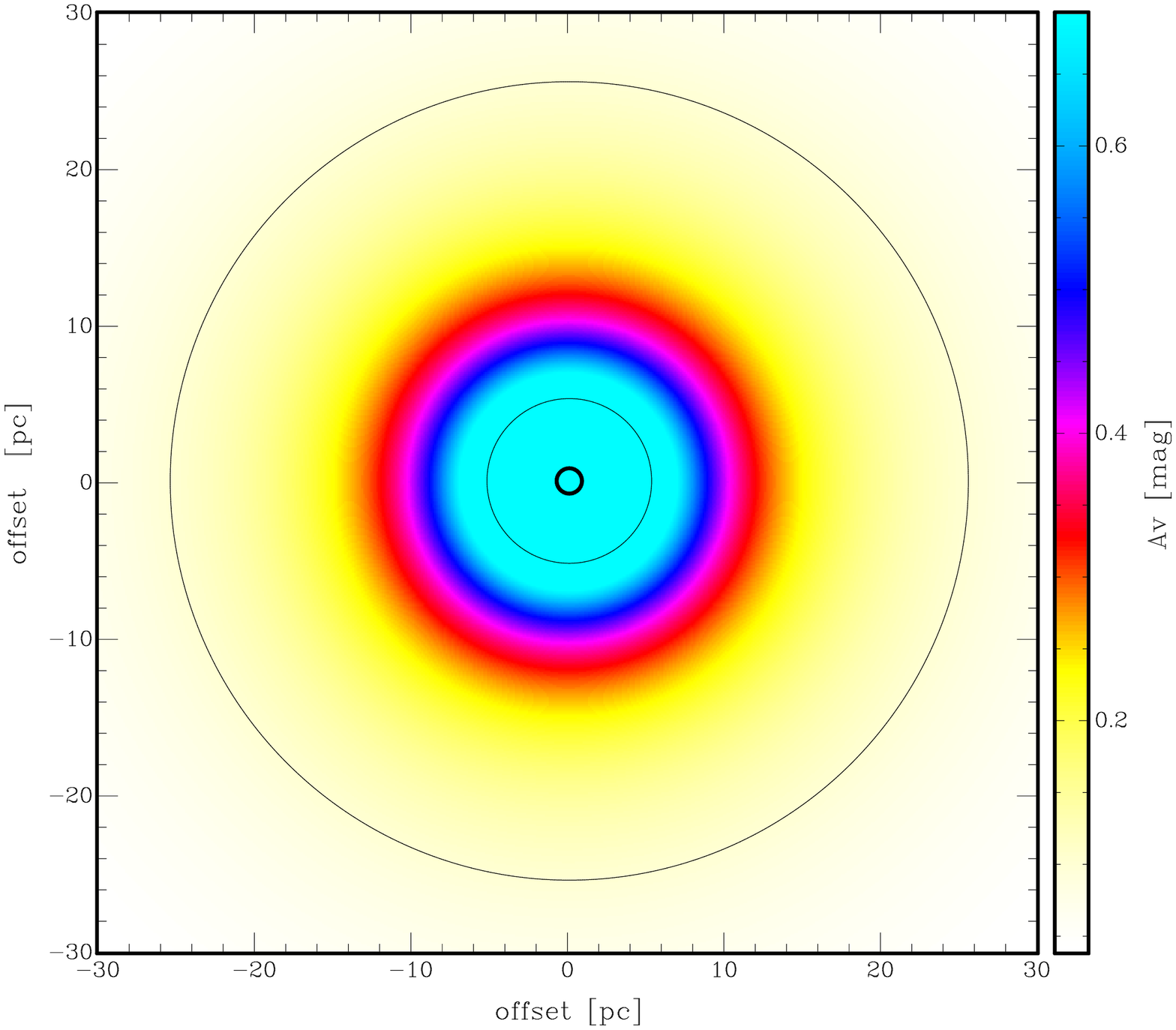} &
\includegraphics[height=0.48\textwidth, angle=-90]{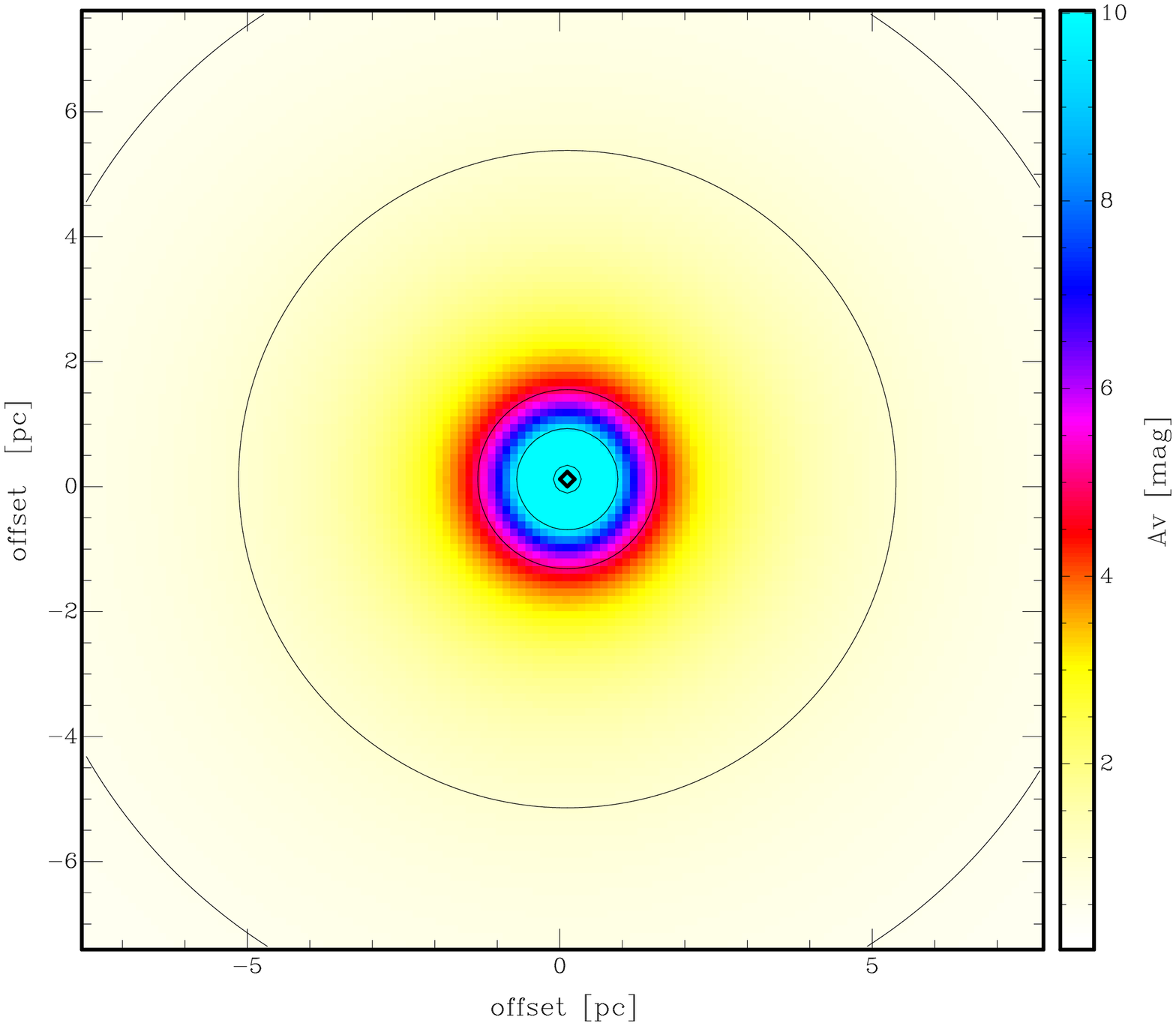}\\
\end{tabular}
\caption{Predicted extinction in the V-band in magnitudes based on the gas profile in the \citet{dercole2008} model (see text for details). The left and right panel show the extinction for the inner 30\,pc and inner 8\,pc, respectively. The colour scale, denoted by the colour bar, accentuates the intermediate extinction regions at each spatial scale. In the left panel, the outer, intermediate and inner black contours show the radius at which the extinction equals 0.1, 1 and 10 mag. In the right panel, the contours show extinctions of 0.5, 1, 5, 10, 50 and 100\,mag.}
\label{fig:Av_map}
\end{figure*}

To determine what the extinction profile might look like, we first projected the 3D volume density profile of the model constructed in $\S$~\ref{sub:infall_model} on to a 2D surface. We then converted the gas column density to a V-band extinction as a function of position using the same conversion as above. The left panel of Figure~\ref{fig:Av_map} shows the extinction map for the inner 30\,pc of the \citet{dercole2008} model. The outer black contour shows that at radii $\lesssim$25\,pc the predicted extinction is $>$0.1\,mag, so should be readily detectable in optical/infrared colour-colour and colour-magnitude diagrams. The right panel of Figure~\ref{fig:Av_map} zooms in to radii of $<$8\,pc. At radii $\sim$5\,pc the extinction reaches 0.5\,mag, and rises to $>$10\,mag at pc-scales.This extinction would have a dramatic affect on optical/infrared observations, causing a pronounced dip in the luminosity profile at small radii. Such a dip should be easily detectable in spatially-resolved observations of young clusters. With multiple filter, high precision, spatially-resolved, optical/infrared photometry, the extinction may be detectable as increased reddening and extreme extinction towards the cluster centre.  In summary, it should be possible to identify any resolved, massive ($\gtrsim10^6$\,M$_\odot$), compact (radii of a few pc and larger) clusters with large central gas reservoirs.\footnote{We note that \citet{dalessandro14} observe multiple stellar populations towards a cluster which they estimate had an initial mass of $\sim2\times10^5$\,M$_\odot$. If this turns out to be common, evidence of centrally-concentrated reddening/extinction may be expected towards clusters much less massive than $10^6$\,M$_\odot$.}

\begin{figure}
\includegraphics[height=0.48\textwidth, angle=-90]{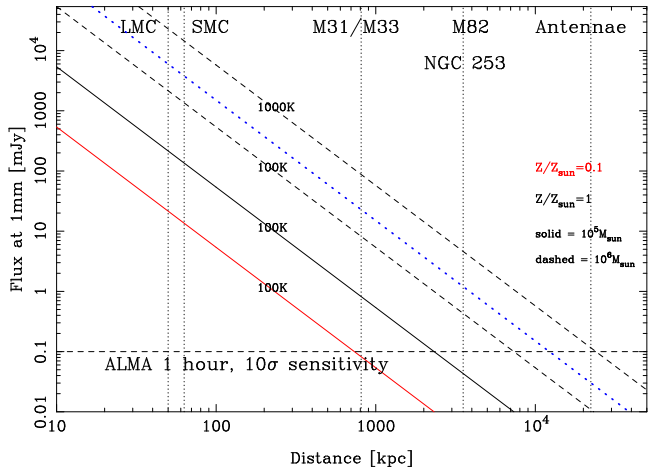}
\caption{Expected flux density at a wavelength of 1mm for gas of a given mass, (dust) temperature and metallicity as a function of distance. The blue dotted line shows the predicted flux based on the \citet{dercole2008} model (see text for details). The horizontal dashed line shows the approximate ALMA 10$\sigma$ sensitivity limit at this wavelength for a one hour observation. The vertical dotted lines show the distances to some nearby galaxies. Large gas reservoirs should be readily detectable towards young massive clusters in many nearby systems.}
\label{fig:dust_emission}
\end{figure}

The dust causing the extinction will also be observable at longer wavelengths in emission. Figure~\ref{fig:dust_emission} shows the expected flux density at a representative wavelength of 1\,mm as a function of distance for gas of different masses, temperatures and metallicities (calculated following \citet{kauffmann08}  using a dust opacity of 0.01\,cm$^2$g$^{-1}$, \citealt{ossenkopf_henning94}) from the idealised GC ($\S$~\ref{sub:toy_model}). The blue dotted line in Figure~\ref{fig:dust_emission} shows the expected 1mm flux from the \citet{dercole2008} model, when extrapolating the density and temperature profiles in $\S$~\ref{sub:infall_model} out to a radius of 250pc (the maximum radius of the model in their Figure 1). The expected large reservoirs of gas should be readily detectable in young massive clusters in galaxies out to several tens of Mpc. 

Finally, we consider what affect variations in the gas distribution would make to the above calculations. The uniform volume filling factor assumed in $\S$\ref{sub:toy_model} represents the lowest possible peak extinction and the lowest possible gas and dust surface brightness. Due to mass conservation, the total amount of gas within the volume considered must remain the same, however it is distributed. Therefore, a higher degree of gas clumping (equivalent to a smaller gas volume filling factor) will tend to make any extinction features more pronounced. In other words, although the area subtended by the extinction will decrease, the actual value of the extinction wherever the gas is confined should increase. 

If the gas distribution is not uniform, the spatial distribution of dense clumps can affect the optical/IR extinction and luminosity profile. For example, although highly improbable, if all the dense clumps happened to lie on the far (near) side of the cluster along our line of sight, then they would sit behind (in front of) the majority of stars. In which case they would block less (more) of the cluster light than if they sat on the opposite side of the cluster along our line of sight. This would result in a less (more) pronounced dip in the optical/IR luminosity profile and less (more) pronounced optical/IR extinction features.

Therefore, combining these effects, if the optical/IR observations can spatially resolve the size of the clumps, and they do not all lie behind the stars, the clumps will be detectable as very pronounced (much higher extinction than calculated in Figure~\ref{fig:glob_extinction}), patchy extinction features. If the observations cannot resolve the typical size scale of the clumps, the expected extinction and gas/dust emission becomes more complicated to determine. The optical/IR luminosity profile will then depend on both the size distribution and location of the clumps. Although certainly interesting, these more detailed calculations are beyond the scope of the current paper.

Unlike the case for optical/IR extinction, the emission from the gas and dust at mm wavelengths will be the same whether the clumps lie at the near or far side of the cluster. If the clumps are resolved, the higher gas volume density will lead to a much higher gas and dust surface brightness (i.e. brightness temperatures), making the gas and dust much easier to detect than shown in Figure~\ref{fig:dust_emission}. If the clumps are not resolved, the observed brightness temperature of the dust and gas emission will be reduced by approximately the ``beam\footnote{The `beam' is effectively the resolution of the observations.} filling factor" -- the total area subtended by all the clumps within the beam, divided by the beam area. 

In summary, if observations can resolve the typical size scale of any gas substructures, the assumption of uniform volume filling factor represents a conservative estimate of the expected peak extinction and minimum expected gas and dust brightness temperature. 

%------------------------------------------------------------
\section{Discussion}
\label{sec:discussion}

Given current instrumental limitations, it would be extremely difficult to detect the presence of gas reservoirs from the luminosity profile or mm continuum emission in young globular clusters at high redshift. However, these features should be readily detectable in clusters of similar stellar mass and density with ages $<$100\,Myr in nearby galaxies with metallicities $>$0.1$Z_\odot$. Neither dips in the optical/infrared luminosity profile nor mm continuum emission have been detected towards any such young massive clusters (YMCs) \citep[e.g.][]{bastian_strader14,cabrera-ziri15}. We now consider potential explanations for this.

A fundamental assumption of the preceding analysis is that the dust properties (grain size distribution, opacities etc) are similar to those observed in the Milky Way and nearby galaxies (e.g. the large and small Magellanic clouds). If the dust properties were different, this would be detectable through a combination of radio and sub-mm wavelength observations with optical/infrared extinction observations. If the grains were predominantly much smaller, one would expect to see copious PAH emission at 8$\mu$m in an environment with such high Lyman-Werner luminosity. To our knowledge, bright PAH emission has not been observed towards the centre of YMCs. Extinction observations are typically not sensitive to dust grains much larger than the observing wavelength, so it may be possible to hide the dust if the grains are all much larger than $\sim$mm. However, in the absence of any plausible physical mechanism to cause all the dust grains to be $>$mm-sized, this seems highly improbable.  We dismiss anomalous grain size distribution variations as an explanation for the lack of dust detected towards young massive clusters.

A second assumption is that the gas-to-dust ratio scales in a similar way with metallicity as observed in the Milky Way, Magellanic clouds and other star-forming galaxies \citep{fisher14}. If the gas-to-dust ratio were anomalously large, it may be possible for clusters to contain large gas reservoirs without any signs of extinction or emission. However, such large gas reservoirs may be detectable in other ways. Once the gas becomes self-gravitating, cools to become molecular and begins forming stars it should be extremely bright in molecular gas tracers such as CO and easily detectable with mm interferometers \citep[see][]{bolatto13}.  The presence of gas with mass of order the stellar mass may also affect the dynamics of the stars, leading to larger-than-expected dynamical mass to light ratios. No molecular line emission or anomalous mass to light ratios have been observed towards YMCs \citep[e.g.][]{cabrera-ziri15}. Again, in the absence of any plausible physical mechanism to cause this, we dismiss anomalous gas-to-dust variations as an explanation  for the lack of gas and dust detected towards young massive clusters.

An alternative explanation is that the gas accumulation and star formation may happen in very short time intervals. This would mean that the chance of catching any individual YMC between 30 and 100\,Myr with detectable amounts of gas would be small. In order for this to happen, a mechanism is required to stop the mass being expelled from AGB stars for a long period of time. The required mass reservoir from AGB stars with the correct abundances would then need to be released and migrate to the cluster centre on a very short timescale, as soon as the required reservoir of pristine material is being accreted. Such a scenario seems highly implausible. Nevertheless, if we take the extreme `burst' scenario, where the gas is visible for 1\,Myr over a 70\,Myr period, we may only expect to see 1/70 YMCs in the appropriate age range with signs of a large gas reservoir. This may provide an explanation for why observations of individual YMCs of sufficient mass do not show any evidence for reddening or extinction \citep[e.g.][]{mccrady05}. However, a direct prediction of this scenario is that observations of large samples of YMCs should recover some sources with extreme extinction/reddening of the luminosity profile. In addition, a consequence of such a large second star formation burst, is that some clusters in this mass and age range should show evidence for signs of extended star formation histories. Recent studies find no such evidence \citep{bastian13b, cabrera-ziri14}.

%------------------------------------------------------------
\section{Conclusions}

If clusters of mass $\gtrsim$10$^6$\,M$_\odot$ and radii of a few pc commonly form a second generation of stars at ages between 30 and 100\,Myr from a large gas reservoir that builds up in the cluster centre, evidence for this gas would be straight forward to observe. The fact that no such observational evidence exists calls into question whether such a mechanism happens regularly for YMCs in galaxies within a few tens of Mpc. Future spatially-resolved optical/infrared and mm-continuum observations of large samples of YMCs are required to determine the fraction of such clusters with evidence of large gas reservoirs and/or recent/ongoing star formation. Such observations would be able to quantitatively test the predictions of GC formation scenarios that require GCs to accrete large gas reservoirs in order to form the second generation of stars. 

\section{Acknowledgements}
I would like to warmly thank the referee, Raffaele Gratton, for constructive feedback that improved the paper. I thank Nate Bastian for stimulating and detailed discussions about the topic of this paper and both Nate Bastian and Diederik Kruijssen for detailed  feedback on the manuscript text. 

%------------------------------------------------------------
\bibliography{gcform_gas}

\end{document}